\documentclass[pre,twocolumn,superscriptaddress,floatfix,showpacs]{revtex4}
\usepackage{graphicx}
\usepackage[latin2]{inputenc}
\usepackage{amsmath}

\newcommand{\pos}{\text{pos}}
\newcommand{\Fext}{F^{\text{ext}}}
\newcommand{\dt}{\Delta t}
\newcommand{\tpdt}{{t+\Delta t}}
\newcommand{\Pt}{\partial_t}
\newcommand{\Px}{\partial_x}
\newcommand{\Q}{q}

\newcommand{\Eq}[1]{Eq.~(\ref{#1})}
\newcommand{\eq}[1]{(\ref{#1})}
\newcommand{\Fig}[1]{Fig.~\ref{#1}}

\begin{document}

\title{Elastic behavior in Contact Dynamics of rigid particles}

\date{\today}

\author{T. Unger}
\affiliation{Department of Theoretical Physics, Budapest University of 
  Technology and Economics, H-1111 Budapest, Hungary}
\affiliation{Institute of Physics, Gerhard-Mercator University
  Duisburg, D-47048 Duisburg, Germany}
\author{L. Brendel}
\affiliation{LMGC, University of Montpellier II, 34095 Montpellier
  cedex, France}
\author{D.E. Wolf}
\affiliation{Institute of Physics, Gerhard-Mercator University
  Duisburg, D-47048 Duisburg, Germany}
\author{J. Kert\'esz}
\affiliation{Department of Theoretical Physics, Budapest University of 
  Technology and Economics, H-1111 Budapest, Hungary}

\begin{abstract}
  The systematic errors due to the practical implementation of the
  Contact Dynamics method for simulation of dense granular media are
  examined. It is shown that, using the usual iterative
  solver to simulate a chain of rigid particles, effective elasticity
  and sound propagation with a finite velocity occur. The
  characteristics of these phenomena are investigated analytically and
  numerically in order to assess the limits of applicability of this
  simulation method and to compare it with soft particle molecular dynamics.
\end{abstract}


% 02.70.Ns Molecular dynamics and particle methods
% 45.70.Cc Static sandpiles; granular compaction

% artifact

\maketitle

\section{Introduction}

In computational physics one can distinguish two different validation
tasks, which have to be solved in order to make simulations a useful
research tool. First one must prove the validity of a simulation model
by comparing its results to laboratory experiments, and second,
equally important, one must assess the systematic errors due to the
practical implementation in order to tell, how precisely the
simulation results reflect the theoretical properties of the simulation model.
In this paper we address the second type of validation problem for the
simulation technique of {\em contact dynamics}, which was
developed about 10 years ago \cite{jean92,moreau94,Jean99} with the
aim to investigate granular media \cite{herrmann98} in the limit of
high rigidity of the 
particles at a high packing density. This method has been successfully
applied and reproduces experiments (see e.g. \cite{Jean97}).
% \bibitem{Jean97} D. Daudon, J. Lanier, M. Jean, A micro mechanical
% comparison between experimental results and numerical simulation of
% a biaxial test on 2D granular material, in: Powders \& Grains 97,
% eds. R. P. Behringer, J.T. Jenkins (Balkema, Rotterdam, 1997)
% pp. 219 -- 222.
However, its systematic errors and the computational effort to keep
them tolerably small have not been investigated in detail before.

This is in marked contrast to other discrete elements methods for
granular media (cf.\ e.g.\ \cite{wolf96}), in particular the
soft particle \emph{molecular dynamics} simulation model, which has
been widely  used since more than 20 years \cite{cundall79}. During
this time  possible pitfalls such as the \emph{detachment
  effect}\cite{luding94} and the \emph{brake failure
  effect}\cite{schafer95} could be discovered, analyzed and hence
avoided. 

Contact dynamics simulations have been applied to study a large
variety of questions in dense granular systems, where excluded volume
interactions and static friction, so called unilateral constraints,
are believed to be essential  
\cite{RadjaiBrendel96,Radjai96,Radjai98,Radjai99,Nouguier00}, but it
should be mentioned that such constraints arise also in other
areas like virtual reality, engineering, especially in robotics, and
operations research, where the numerical treatments are similar
\cite{Lynch94,Stewart96}. 

In a system of perfectly rigid particles the sound velocity would be
infinite. This is in principle borne out by the contact dynamics
simulation model. However, its practical implementation will normally
give rise to sound-waves in the granular material, as we are going to
show in the following, even if
each single collision is modeled as being perfectly inelastic. Our aim
is to elucidate this artifact after introducing briefly the principles
of this method.

\section{The Contact Dynamics Method}

\label{CD-sec}
% Computer Simulation

% The method was developed by M.~Jean and J.J.~Moreau and proved to be very
% useful in applications, where dry friction and hard particles are
% present. It provides an interesting alternative to the popular Molecular
% Dynamics (MD) in the context of the granular media. We would like to point
% out that the two methods have basically different approach to calculating
% the forces between the grains.

First, let us point out the basic difference between molecular
dynamics (MD) on one side and contact dynamics 
(CD) on the other. Both have in common the integration of Newton's
equation of motion where the occurring forces are due to external
fields (gravity) or are -- more important -- \emph{contact forces},
i.e.\ caused by contacts between particles or their contacts with
confining walls.

The spirit of MD is to calculate the contact forces according to
their \emph{cause}, i.e.\ the (usually microscopic) deformation of
the contact region and the involved velocities. Since the full treatment
of every particle as a deformable body would render the simulation of
a large number of particles exceedingly time consuming, a lot of
models exist, how to replace the deformation by the local overlap
\cite{schafer96}, the latter being a 
virtual quantity obtained from the undeformed shapes.

The principles of CD are different: Here the contact forces are
calculated by virtue of their \emph{effect},
% \emph{action}\footnote{This does not
% refer to the physical quantity of dimension energy times time.},
which is to fulfill certain constraints. Typically, such a
constraint is the volume exclusion of the particles or the absence of
sliding due to static friction. As can be seen immediately, this problem cannot
be solved locally: In a cluster of particles where many contacts are
simultaneously present, the force at one contact depends on adjacent
contacts and so on. In that case the aim is to find a global
force-network, which is consistent with the constraints at
all contacts. The method to carry out this calculation is often called
the \emph{solver} in this context, which is commonly an iterative
scheme, as the one we describe in the following section.
%In order to make our points very clear, we present them for a one
%dimensional example, though they can be generalized to higher
%dimensions as well.
In order to make our points very clear, we perform an analytical
investigation for a one dimensional example but can prove the existence
of the discovered effect in two dimensions as well.

\subsection{The dynamical equations}

As CD is designed to obey excluded volume constraints exactly,
particle collisions lead to discontinuous velocity changes
(``shocks''), i.e. to nonsmooth mechanics \cite{moreau88}. 
Therefore, higher order terms than
employed in the Euler integration scheme are of no use. For the
$i$th particle's positions $x_i$, this reads
% For the integration of the equations of motion an implicit Euler method is
% applied. It is not worth using higher order schemes since
% one starts with the idea of perfectly hard particles and non-smooth
% (discontinuous) velocities are expected due to shocks.
% In one dimension our discrete equations have the form:
%the following first-order approximation is applied:
\begin{equation}
  x_i(t+\dt) = x_i(t)+v_i(t+\dt) \dt \ ,
                                % \null_{\text{pos-update}}
  \label{pos-update}
\end{equation}
and correspondingly for its velocity $v_i$ we have
\begin{equation}
  v_i(t+\dt) = v_i(t) + \frac{F_i(t+\dt)}{m} \dt \ ,
                                % \null_{\text{veloc-update}}
  \label{veloc-update}
\end{equation}
where $F_i$ is the total force acting on the particle, $m$ its mass
and $\dt$ is the time step.

A remark about the Euler scheme \eq{veloc-update} being
\emph{implicit} is in order: Whereas in conventional MD the choice of
evaluating the force for the previous or for the new
configuration (i.e.\ at $t$ or $t+\dt$ respectively) is merely a
matter of stability of the integration scheme (cf.\
e.g.~\cite{nr_stiff}), the constraints in CD can only be imposed on
the \emph{new} configuration. In this sense, the integration scheme of CD is
inevitably of implicit type. (Taking into account the configurational
change during a time step consistently (\emph{fully implicit}
integration \cite{Jean99}) leads to difficulties which are analogous
to implicite schemes in MD, when the forces have to be evaluated for
the yet unknown new configuration. In one dimension, though, these
difficulties do not arise.)

\subsubsection{One contact}

\begin{figure}
  \begin{center}
    \includegraphics[scale=0.7]{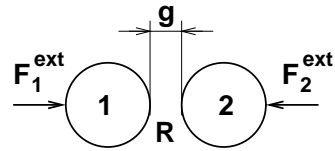}
  \end{center}
  \caption{A pair of particles\label{fig-one-contact}}
\end{figure}

We now turn to the force on the $i$'th particle, $F_i$, occurring in
\Eq{veloc-update}: In order to determine it  one has to
know the contact forces between the grains. In CD they are calculated
from the condition that the constraints must not be
violated. In one dimension and if one disregards rotations, this is
simply the excluded volume contraint. To give a specific example, let
us consider  two particles with equal masses $m$ subjected to constant
external forces (\Fig{fig-one-contact}). A contact force $R$
(which is the reaction force due to the constraint) is active only, if
interpenetration needs to be prevented. Otherwise, i.e.\ if the gap $g$ would
remain non-negative (no overlap) anyway, it takes on its minimal value,
$R=0$ (this is expressed in the CD literature as \emph{Signorini's
  condition}).
Without the repulsion $R$ between the particles the gap at the end of
a time step would be given by  
\begin{equation}
g' = g_t + (v_{2,t} - v_{1,t})\dt - \frac{\Fext_1+\Fext_2}{m} \dt^2
\label{g_prime}
\end{equation}
according to \eq{pos-update} and \eq{veloc-update}. However, the excluded
volume constraint requires that $g_\tpdt = \max\{g',0 \}$.
%\begin{equation}
%g_\tpdt = \left{ \begin{array}{l} 
%                 g' \; {\rm if} \; g'>0 \\ 
%                 0  \; {\rm if} \; g'\leq 0
%                 \end{array}\right. .
%\end{equation}
In order that this results from \eq{pos-update} and \eq{veloc-update},
the contact force $R_\tpdt = \max\{R',0 \}$ must be taken into
account, where
\begin{eqnarray}
R' &=& - \frac{m g'}{2\dt^2} \\
   &=& \frac{m}{2 \dt} \left(\frac{-g_t}{\dt} +
    v_{1,t} - v_{2,t}\right) + \frac{\Fext_1+\Fext_2}{2}              
  \label{gap-force1}
\end{eqnarray}
%\begin{equation}
%R_\tpdt = \left{ \begin{array}{l} 
%                 0  \; {\rm if} \; g'>0 \\ 
%                 - m g'/2\dt^2  \; {\rm if} \; g'\leq 0
%                 \end{array}\right. .
%  \label{gap-force1}
%\end{equation}
Note that this scheme corresponds to a completely inelastic collision
(i.e.\ with the so-called \emph{restitution coefficient} being zero),
which is accomplished in two time steps: At first the gap closes, in
the next step also the relative velocity vanishes. (Finite restitution
coefficients can also be incorporated into this algorithm
\cite{Radjai98a}.)

% \bibitem{radjai98a} F. Radjai, Multicontact Dynamics, in: Physics of
% Dry Granular Media, eds. H.J. Herrmann, J.-P. Hovi, S. Luding
% (Kluwer Academic Publishers, Dordrecht, 1998) pp.305 --311.

The above determination of the contact force has a drawback, though:
If $g_t<0$ occurs due to a previous inaccuracy, then the
\emph{elimination} of this overlap is accompanied by a surplus of
kinetic energy.  Therefore, mostly the \emph{quasi-inelastic
  shock} formula \cite{Jean99}
\begin{equation}
 R'=\frac{m}{2 \dt} \left(
   \frac{- g_t^{\text{pos}}}{\dt} +
   v_{1,t} - v_{2,t}
 \right) + \frac{\Fext_1+\Fext_2}{2}
                               % \null_{\text{force-update}}
 \label{force-update}
\end{equation} 
is used instead of (\ref{gap-force1}), where $g_t^{\pos}=\max\{0,
g_t\}$. That means, negative gaps are treated differently from
\Eq{gap-force1} in such a way that an already existing overlap is
not eliminated but only its further growth is inhibited. Hence
the inelastic shock law \eq{force-update} is in a way even ``more
inelastic'' than the original law \eq{gap-force1}, because it avoids
overlap correcting impulses which could destroy stable equilibrium
states.

\subsubsection{Many contacts}

\label{sec:iteration}

\begin{figure}[b]
  \begin{center}
    \includegraphics[scale=0.7]{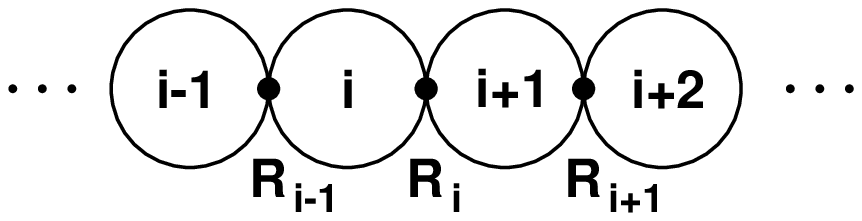}
  \end{center}
  \caption{A multi-contact situation in an 1D array. External forces
    act only on particles far away from those shown. Each particle is
    subjected only to the contact forces of the adjacent ones.
  \label{fig-multi-contact}} 
\end{figure} 
We now address the question how to solve the problem of the constraint
forces if we consider not only one, but many contacts at the same
time. \Fig{fig-multi-contact} shows such a system, where $R_i$
denotes the contact force between the particles $i$ and $i+1$, and for
the sake of simplicity we have no external force acting on them.
Furthermore, we will concentrate on the situation where the
neighboring pairs are permanently in contact (i.e.\ $g^{\pos}_{i,t}=0$)
which corresponds to compressed dense packings.

For the $i$th contact in this setup, $R_{i-1}$ and $R_{i+1}$ play the
role of the external forces in \Eq{force-update}, though they
are not constant, their values are not known for the next time step:
\begin{equation}
  R_{i,\tpdt}=\frac{m}{2 \dt} 
  \left(
    %%\frac{g_{i,t}^{\pos}}{\dt} + 
    v_{i,t} - v_{i+1,t}\right) + 
  \frac{R_{i-1,\tpdt}+R_{i+1,\tpdt}}{2} \ .
                                % \null_{\text{force-update2}}
  \label{force-update2}
\end{equation} 
Obviously the contact force is coupled to the neighboring contacts and
through those to further contacts. In a similar way in higher
dimensions, large numbers of contacts are coupled within clusters
defined by the contact network. Hence, the determination of the proper
reaction forces becomes a global problem.

A standard way in CD is to solve this problem iteratively: In one
iteration step we calculate the forces according to the constraint
conditions pretending that the corresponding neighboring contacts
already exhibit the right forces. In that way the process traverses
the list of contacts many times until satisfactory convergence is
reached (for the question of the convergence cf.\ the works
\cite{convergence,existence}).

For our one-dimensional chain of particles, this means that
\Eq{force-update2} simply gets the meaning of an
\emph{assignment} of the right hand side to the left hand side, and
% \begin{equation}
%   R_{i}^{\act}:=\frac{m}{2 \dt} 
%   \left(
%     %% \frac{g_{i,t}^{\pos}}{\dt} + 
%     v_{i,t} - v_{i+1,t}
%   \right) + 
%   \frac{R_{i-1}^{\act}+R_{i+1}^{\act}}{2} \ ,
%   \null_{\text{force-update-iter}}
%   \label{force-update-iter}
% \end{equation} 
% where $R_i^{\act}$ is the actual value of the contact force, that changes
% gradually during the iteration from the value $R_{i,t}$ to $R_{i,\tpdt}$.
one iteration step consists of applying this assignment sequentially
once to each contact. The order of this sequence is preferably random
(with the pattern changing for every sweep), in order not to create
any bias in the information spreading\footnote{We call
  this procedure ``random sweep'' as opposed to ``random sequential''
  where the choice of the contact to be updated would be random
  \emph{and independent} of the previous choices.}. With each sweep
a globally consistent solution is approached, until finally a chosen
convergence criterion is fulfilled.
% With the iteration loop we get more and more close to a consistent
% force-system. The conception is that if the forces are accurate
% enough, the iteration is stopped, we set $R_{i,\tpdt}=R_i^{\act}$ and
% with the help of the new forces the integration of the Newton
% equations can be proceeded.

In the next section we shall show, what kind of consequences the
local update scheme presented here has on large time and length
scales.

\section{The large scale description}

\label{contin-sec}

In order to analyze the coarse grained behavior of the microscopic
equations derived in the last section we can also regard them as 
the discretized form of a continuum description, making a
treatment in terms of partial differential equations (PDE) possible. In
order to obtain the corresponding PDEs, we consider the particle index
$i$ as space variable $x$ and replace the differences of consecutive
quantities by derivatives, the error term for the first and second
order derivatives being of first and second order, respectively. E.g.\ 
$v_\tpdt - v_t \to \dt\,\Pt v$ and $R_{i+1}+R_{i-1}-2 R_i \to d^2
\Px^2 R$, where $d$ is the particle diameter.

\subsection{The relaxation of the contact forces}

While the continuum versions of the updates \eq{pos-update} and
\eq{veloc-update} can be obtained straight forwardly, the force
change \eq{force-update2} lacks a time variable, for during the
force-iteration, being just a calculation, no physical time passes.
Hence, to be able to describe this force-development as well, let us
introduce a fictitious time $t^*$ with time interval $\dt^*$ for one
iteration-step. With this, the continuum version of
\Eq{force-update2} reads:
\begin{align}
  \label{contin-forceiter}
  \partial_{t^*} R &= D \Px^2 R - \beta \Px v\\
                                %\null_{\text{contin-forceiter}}
  \text{with}\quad
  D &= \Q \frac{d^2}{\dt^*}\\
  \beta &= \Q \frac{md}{\dt\dt^*}\\
  \text{and}\quad
  \Q &= \frac{1}{2}
\end{align}
This analytic form clearly reveals the nature of the iteration loop:
The reaction forces relax towards the solution in a diffusive way.
(Note that the $\Px v$ term is constant in $t^*$, it only depends on
$x$.)

The introduction of the constant $\Q$ reflects a subtlety regarding
the sequential character of the update discussed in the appendix. In fact,
since 
the PDE \eq{contin-forceiter} describes the change of the whole
field $R(x)$ \emph{at once}, given its actual value at time $t^*$, it
corresponds to a \emph{parallel update} (in the sense that the right
hand side of \Eq{force-update2} always employs the values $R_i$
from the \emph{beginning} of the iteration sweep, not the freshly
updated ones). In appendix \ref{squential-appendix} we shall show,
though, that a random sweep update instead of a parallel one only
renormalizes the value of $\Q$ to about $0.8$ while leaving the form
of the PDE untouched.

\subsection{Sound waves}

\label{sound-sec}

To connect the velocity update, whose continuum version follows
immediately from \Eq{veloc-update} as
\begin{equation}
  \Pt v = -\frac{d}{m} \Px R \ ,
                                % \null_{\text{contin-veloc-update}}
  \label{contin-veloc-update}
\end{equation}
to the force update, we must relate the ``iteration time'' $t^*$ to
the physical time $t$. Although, depending on the convergence
criterion, there can be in principle a varying number of iterations
during one physical time step $\dt$, we assume for simplicity this
number $N_I$ being fixed. (Actually, in practice this crude
``criterion'' is sometimes applied.)

Hence, with $\dt=N_I\dt^*$, we can express everything in terms of the
physical time:
\begin{align}
  \label{contin-forceiter2}
  \partial_t R &= D \Px^2 R - \beta \Px v \ ,\\
                                % \null_{\text{contin-forceiter2}}
  \text{and}\quad
  D &= \Q N_I  \frac{d^2}{\dt} \\
                                % \null_{\text{D}}
  \beta &= \Q N_I \frac{md}{\dt^2}
\end{align}
With the equations \eq{contin-veloc-update} and
\eq{contin-forceiter2} we obtained two coupled PDEs. We can combine
them to arrive at a wave equation with an additional
% diffusion type
damping term:
\begin{equation}
  \Pt^2 R= c^2 \Px^2 R + \Pt\left( D \Px^2 R \right)
                                % \null_{\text{wave-eq}}
  \label{wave-eq}
\end{equation}
The sound velocity appearing is of finite value:
\begin{equation}
  c = \sqrt{\Q N_I} \frac{d}{\dt}
                                % \null_{\text{soundvelocity}}
  \label{soundvelocity}
\end{equation}

This equation indicates that the CD simulation of the particle chain,
as presented in the previous section, can lead to sound
propagation like in an elastic medium,
% We found those astonishing
% sound waves also in numerical experiments
which however contradicts the conception of perfect rigidity. The
constraint conditions applied at the contacts should in principle
prohibit overlaps, i.e.\ prohibit elastic deformation of the
grains. It can be seen that this deviation from the perfect rigidity
enters at the force relaxation: A finite number of iterations means a
finite range for the information spreading and thus yields systematic
errors in the calculated reaction forces. As a consequence, the finite
$N_I$ involves soft particles and a finite sound velocity $c\sim
\sqrt{N_I}$.  Note that in the limit of an infinite $N_I$, the exact
value of the forces is reached, which corresponds to the case
$c\to\infty$, as it should be for rigid particles.

\subsubsection{Dispersion}

Performing a Fourier transformation on \Eq{wave-eq}, one
obtains the properties of the different wave modes. The oscillation
frequency $\omega$ of the wave number $k$ is
\begin{equation}
  \omega\left(k\right)=k \sqrt{c^2-\frac{D^2 k^2}{4}} \ .
                                % \null_{\text{dispersion}}
  \label{dispersion}
\end{equation}
That means, $\omega\left(k\right)$ becomes zero at a critical wave
number
\begin{equation}
  k_c = \frac{2 c}{D} \sim \frac{1}{\sqrt{N_I}}
                                % \null_{\text{k\_c}}
  \label{k_c}
  ~,
\end{equation}
and waves with $k$ larger than $k_c$ (short wave lengths) are over-damped. The
damping time $\tau(k)$ for the oscillating modes is given by:
\begin{equation}
  \tau(k)= \frac{2}{D k^2}
                                % \null_{\text{tau}}
  \label{tau}
\end{equation}

We derived the dispersion relation \eq{dispersion} in the continuum
limit which is a good approximation for small wave numbers, but not
near to the border of the Brillouin zone ($k_{\text{Br}}=2\pi/d$),
where the effect of the spatial discreteness can be strong. However,
increasing the number of the iterations sufficiently, $k_c$ becomes
small compared to $k_{\text{Br}}$.  Actually, for $N_I \ge 10$ the
formula \eq{dispersion} works well not only for small wave numbers but
for all the oscillating modes, as could be verified numerically.

\subsubsection{Numerical confirmation}

\begin{figure}[b]
  \begin{center}
    \includegraphics[scale=0.5,angle=0]{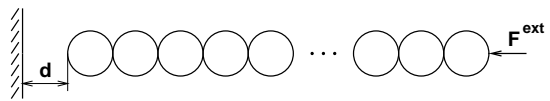}
  \end{center}
  \caption{The initial configuration of the numerical experiment for testing
    the properties of the sound waves.\label{fig-oscill-start}}
\end{figure}

In order to confirm the results of this section, we performed the
following numerical experiment: The starting configuration of the
simulation consists of an array of $50$ discs and an immobile wall,
the geometry can be seen in \Fig{fig-oscill-start}. Initially the gap
between the wall and the leftmost particle is one disc diameter ($d$),
the gap between the particles is zero and the array has zero velocity.
Starting from $t=0$ a constant external force ($\Fext$) is acting on
the rightmost particle which accelerates the array towards the wall
(only horizontal motion takes place). As
simulation parameters we chose $N_I=40$ and $\Fext= 0.05\:d m
\dt^{-2}$.

\begin{figure}
  \begin{center}
    \includegraphics[scale=0.27,angle=-90]{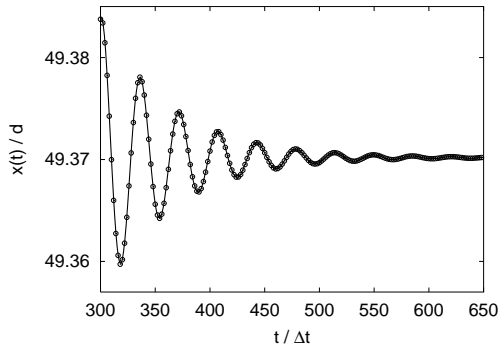}
  \end{center}
  \caption{%
    Damped oscillation in a Contact Dynamics simulation. The dots
    indicate the measured data: the position of the rightmost particle versus
    time (for details see the text). The line is an exponentially damped sinus
    function, where the frequency and the damping time is provided by our
    continuum model. \label{fig-oscill}}
\end{figure}

The collision with the wall induces a relative motion of the grains
and generates sound waves in the array. After a transient period the
grains remain permanently in contact (the whole array is pressed
against the wall by $\Fext$). Since the different wave modes have
different relaxation time, after a while only the largest wave length
mode survives. This wave length is four times the system size because
the wall represents a fixed boundary while the right side is free.
Since the wave length is given, the oscillating frequency and the
damping time can be calculated from \Eq{dispersion} and \Eq{tau},
respectively. For comparison with the simulation we measured the
motion of the rightmost particle. The expected motion is a damped
oscillation
\begin{equation}
  x(t) = x_0+
  A \exp\left(-t/\tau\right)\sin\left(\omega t+\phi\right)
  ~,
\end{equation}
where the offset $x_0$, the amplitude $A$ and the phase shift $\phi$
have to be fitted (in contrast to $\omega$ and $\tau$) for a
comparison. In \Fig{fig-oscill} the measured data (dots) and the
fitted curve can be seen. It shows that the simulation is in good
agreement with our continuum description.

\subsection{Global Elasticity}
\label{elastic-sec}

It is instructive to compare our test-system to its simplest MD
counterpart where the contact forces depend linearly on the local
kinematic variables, i.e.\ the so called \emph{linear spring/dashpot model}
\begin{equation}
  R_i= -\kappa \left( x_{i+1}-x_i-d \right) 
       -\gamma \left( v_{i+1} - v_i \right)
  \label{lin-spring}
\end{equation}
with the spring stiffness $\kappa$ and the damping coefficient
$\gamma$. Employing again the updates \eq{pos-update} and
\eq{veloc-update} for the positions and velocities, respectively,
the continuum limit yields the same type of PDE as
\Eq{contin-forceiter2} with its coefficients being inherited
from \Eq{lin-spring}:
\begin{equation}
  \label{spring-R}
  \partial_t R = \frac{\gamma d^2}{m} \Px^2 R - \kappa d \Px v
\end{equation}
This allows us to relate the physical MD model parameters to the
``technical'' CD parameters:
\begin{align}
  \label{kappa}
  \kappa & = \Q m \frac{N_I}{\dt^2}
  \intertext{and}
  \label{gamma}
  \gamma & = \Q m \frac{N_I}{\dt}
\end{align}
This equivalence shows that on large scales the CD chain should behave
identical to its MD counterpart, e.g.\ it will exhibit a global
shrinkage proportional to an external compressive load. Note that a
real congruence can be expected only for the collective behavior but
not on the level of the contacts. In the CD method, as explained
above, the contact forces are not related to the overlaps, which must
merely be regarded as due to the incompleteness of the
force-calculation and in fact are stochastic quantities because of our
random update procedure.  Only on scales larger than the grain size,
where the fluctuations of these local ``deformations'' are averaged
out, the behavior can be smooth like in an elastic medium, as is shown
in \Fig{fig-oscill}.

\medskip

In sections \ref{sound-sec} and \ref{elastic-sec}, our calculation was
based on the assumption of a constant number of iterations for every
time step, and due to this premise the analytical treatment became
simple and directly comparable to the corresponding simulation. We
should keep in mind, though, that the application of a convergence
criterion involves a fluctuating $N_I$ (i.e.\ it can vary from time
step to time step), and therefore steps with a different ``stiffness''
are mixed during the integration of motion. Consequently, the behavior
of the CD method is more complex in detail, but qualitatively the
results for the constant $N_I$ remain relevant also here. (E.g.\ the
mechanism resulting in soft particles or the way how shock-waves can
arise with finite velocity.)

% After all, the diffusion like
% calculation of the forces is unchanged (i.e.\ 
% eq.~(\ref{contin-forceiter}) stays the same) whether we use a
% fluctuating $\I$ or not.

\section{2D simulation}

After the analysis of the regular 1D system, the important question
arises whether the behavior is similar in higher dimensions and
less regular systems. Hence, we performed CD simulations with
two-dimensional random packings of discs and observed the same
``elastic'' waves (even transversal modes were found).

\begin{figure}
  \begin{center}
    \includegraphics[scale=0.5]{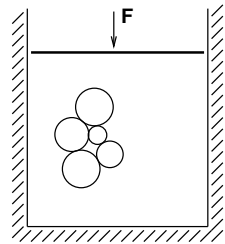}
  \end{center}
  \caption{The setup of a numerical experiment in two dimensions. A dense
    packing of 1000 discs is prepared in a container via compressing
    the system by means of the mobile upper wall.
    \label{fig-geometry2D}}
\end{figure}

The simulation presented here consists of $1000$ discs with
radii distributed uniformly between $r_{\text{min}}$ and
$r_{\text{max}}=2 r_{\text{min}}$, the mass of each disc being
proportional to its area. \Fig{fig-geometry2D} shows the geometry: The
base and the two side-walls are fixed while the upper piston is mobile.
Starting from a loose state, we compressed the system and waited until
the packing reached an equilibrium state (the compression force $F$
applied on the piston was kept constant). The simulation was carried
out without gravity and with a Coulomb friction coefficient of $0.05$
for all the disc-disc and disc-wall contacts (cf. \cite{kadau02}).
%
% \bibitem{kadau02} D. Kadau, G. Bartels, L. Brendel, D.E. Wolf,
% Contact Dynamics Simulations of Compactiong Cohesive Granular
% Systems, Computer Physics Communications, in print (2002). 

\begin{figure}
  \begin{center}
    \includegraphics[scale=0.27,angle=-90]{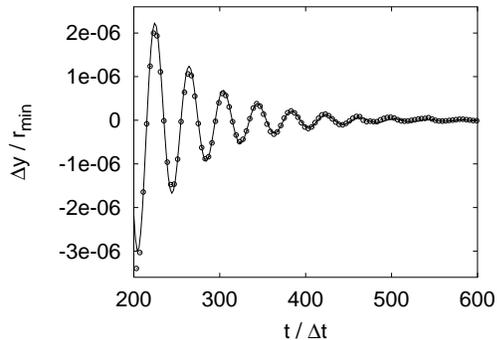}
  \end{center}
  \caption{Oscillations in a 2D simulation are similar to the 1D case. Here the
    sound waves are generated in a random dense packing of discs. The
    dots are the measured position of the upper wall versus time (see
    \Fig{fig-geometry2D}), while the curve is a fitted exponentially
    damped sinus function. \label{fig-oscill2D}}
\end{figure}

After the packing was relaxed completely, we generated sound waves by
increasing the compression force abruptly to $F+\Delta F$. After a
transient period only one standing wave mode survives (both the
wavenumber vector and the collective motion are vertical), where the
piston, representing a free boundary, oscillates with a relatively
large amplitude. We measured the vertical position of the piston versus
time and found that the data can be fitted by an
exponentially damped sine function (\Fig{fig-oscill2D}).
Here, in contrast to the 1D case, also $\omega$ and $\tau$ are fit
parameters, since, due to the different geometry, the
values \eq{dispersion} and \eq{tau} cannot be adopted, but, because of
the random structure of the system, a more complex treatment is
required for a quantitative description. However, we checked the most
important relation, namely that the scaling properties of $\omega$ and
$\tau$ remain valid also for the 2D random system, that is $\omega
\sim \sqrt{N_I}$ and $\tau \sim N_I^{-1}$, which means that the
artificial visco-elasticity of the particles depends on the number of
the iterations in the same manner as we showed for the 1D chain.

\section{Discussion}

The artificial elasticity found in CD simulations was analyzed. We
showed that the systematic errors of the force calculation can lead to
a collective elastic behavior, even though single contacts are assumed
perfectly rigid and perfectly inelastic. For the 1D chain of
particles, we could, starting from the ``microscopic'' laws, reproduce
the numerical results analytically, including the dependence of the
effective stiffness and viscous dissipation of the contacts on the
computational parameters ($N_I, \Delta t$).

Besides elucidating the origin of the elastic behaviour, the coarse grained
description reveals important characteristics of the CD
method, which were less obvious on the discrete level. We saw that
using the iterative solver, the proper contact forces are approached
in a diffusion like manner which is a crucial information concerning
the computational time. The conception of perfectly rigid particles
requires that the calculated forces are consistent even for contacts
far from each other. Therefore, the ``diffusion length'' $\sqrt{D\dt}$
must be larger than the linear system size $L$, which defines a lower
boundary for the number of iterations of the order $(L/d)^2$. The same
condition is obtained if we want to avoid other consequences of the
effective softness. E.g.\ if we want the sound to travel a larger
distance than $L$ during one time step (i.e.\ $c>L/\Delta t$) or if we
would like all possible wave modes to be overdamped (i.e.\ with
$\lambda_c >L$ no oscillations). All these cases are equivalent, one
is forced to apply a relatively large number of iterations: $N_I \sim
L^2$. Going further on this line, the scaling with respect to the
number of particles $n$ can be determined: One step of the iteration
consists of as many force-updates as there are contacts, which is
proportional to $n$. Therefore, the computational effort of one time
step scales with the particle number like $n N_I$, which is $\sim n^2$
in 2D or $\sim n^{5/3}$ in 3D. Therefore a large CD-simulation is
computationally more costly than MD, where the computational effort
scales like $n$. This is the price for simulating rigid particles
without getting elasticity artifacts, which cannot be done with MD.

To avoid this super-linear scaling when dealing with large systems, we
can also accept the finite stiffness by keeping $N_I$ constant
independently of $n$. Then, besides gaining a running time of order $n$,
of course, elastic deformations and sound
waves can arise with an increasing number of the particles, and
consequently they have to be monitored. We want to mention the idea,
though, that in certain situations advantage can be taken of the
artifact. E.g.\ when being applied deliberately, Coulombian friction
can be combined with global elasticity easily;
this way considerable computational time could be saved and even 
better performance than MD could be achieved.

% Suppose one is interested in the large scale behaviour of 
% particles with elastic constant $\kappa$ and
% restitution coefficient $e_n$. A simulation with $n$ particles,
% which covers a real time interval $T$ requires a computational
% effort in MD, which is approximately $nT/\dt \approx nT 100
% \sqrt{\kappa/m}$. In CD the computational effort is $n N_I T/\dt
% \approx nT (1-e_n) \sqrt{\kappa/m}$, where
% $(1-e_n)=\gamma/\sqrt{m\kappa}$ has been identified with
% $\sqrt{N_I}$ according to the identifications (\ref{}) and
% (\ref{}). This shows that it depends on the proportionality factors
% which method requires less computation time. The
% proportionality factor for CD can be up to 100 times larger than
% the one for MD for CD to be less time consuming!

\section{Acknowledgments}

This work was supported by DAAD and Federal Mogul Technology GmbH.
Partial support by OTKA T029985, T035028 is acknowledged.

\appendix

\section{$\Q$ for the random sweep update}

\label{squential-appendix}

To find the proper value of the constant $\Q$ appearing in
sec.~\ref{contin-sec}, two more things have to be taken into account:
Firstly the sequential type of the force update, and secondly that the
order is random. The latter results in a stochastic force relaxation,
i.e.\ the change of $R_i$ in one iteration sweep is a stochastic
variable. In order to obtain a similar equation as
\eq{contin-forceiter} we shall determine the average value $\langle
\Delta R_i \rangle$, where the average for site $i$ is meant to be
taken over all possible update sequences.

Before going any further, let us introduce a few notations:
\begin{itemize}
\item For $N$ being the total number of contacts, the mapping $u :
  \{1,\ldots,N\} \to \{1,\ldots,N\}$ denotes the order of the update
  sequence; i.e.\ if the contact labeled $i$ is updated before $j$,
  then $u_i < u_j$.
  
\item Throughout this appendix, the notation $R_i$ means the value
  from the beginning of the iteration sweep, the value at the end is
  $R_i+\Delta R_i$.

\item We define $\delta R_i$ as the change according to a parallel
  update (cf.\ \Eq{force-update2}), i.e.\ 
  \begin{align}
    \nonumber
    \delta R_i
    &= \frac{m}{2 \dt} \left(v_{i} - v_{i+1}\right)\\
    \label{DRnull}
    &+ \frac{R_{i-1}+R_{i+1}-2 R_i}{2} \ ,
  \end{align}
  as opposed to the \emph{total} change $\Delta R_i$.
\end{itemize}
Given that, it can easily be seen how $\Delta R_i$ depends on the
update order. If e.g.\ the site $i$ is updated earlier than its
neighbors (i.e.\ $u_{i-1} > u_i < u_{i+1}$), then $\Delta R_i =
\delta R_i$, but in the case $u_{i-1} > u_i > u_{i+1} < u_{i+2}$, we
get $\Delta R_i=\delta R_i + \delta R_{i+1}/2$ (because for contact
$i$, \Eq{force-update2} employs the already updated force at
contact $i+1$). Similarly (but with less probability) even very far
contacts can contribute to $\Delta R_i$, which can be summarized in
the following way:
\begin{equation}
  \label{eq:DR_avg}
  \langle \Delta R_i \rangle= \delta R_i + \sum_r \frac{1}{2^r}
  \left( p_r \delta R_{i-r} + p_r \delta R_{i+r} \right)
\end{equation}
Here, $p_r$ is the probability that $\Delta R_i$ contains information
from the update of a contact at distance $r$, that is $p_r =
P(u_i>u_{i+1}>\ldots >u_{i+r})$. (This definition holds true for
contributions from contacts with labels higher than $i$, but due to
left-right symmetry the same value is inevitably obtained for the
corresponding lower ones.)

The value of $p_r$ can be obtained from the following combinatorial
consideration: Given an index $i$ and a distance $r$, we can classify
the set of all update orders into groups such that the sequences in
one group differ only in the permutations of the elements $u_j$, $i
\le j \le i+r$.  Such a group contains $(r+1)!$ sequences, but only
one of them satisfies the condition
$u_i>u_{i+1}>u_{i+2}>\ldots>u_{i+r}$. Since all update sequences are
equally probable, the value of $p_r$ is equal to $1/(r+1)!$.

The factor $(2^r (r+1)!)^{-1}$, relating the contacts $i$ and $i+r$,
decays faster than exponentially; already for $r=8$, it drops below
$10^{-6}$. Therefore, the sum in \Eq{eq:DR_avg} reaches only
the immediate vicinity of contact $i$, such that, for our large
wavelength considerations, the approximation $\delta
R_{i+r}\approx\delta R_i$ can be applied. This allows us to calculate
the average change of the contact force:
\begin{align}
  \nonumber
  \langle \Delta R_i \rangle &= 
  \delta R_i \left(1+2 \sum_{r=1}^{\infty} \frac{1}{2^r (r+1)!} \right)\\
  &= \delta R_i \left(4 \sqrt{e} - 5\right)\ . \label{average-dR}
\end{align}

Thus, it is shown that the random sweep results in a larger change of
$R_i$ than the parallel update. \Eq{average-dR} provides also the
sought value of the parameter $\Q$ as
\begin{equation}
  \Q=\frac{4 \sqrt{e} - 5}{2} \approx 0.797
  ~,
\end{equation}
which completes the continuum description given in sec.~\ref{contin-sec}.

\bibliography{granmat}

\end{document}